\documentclass{article}
\pdfoutput=1
\usepackage{arxiv}
\usepackage[T1]{fontenc}   
\usepackage{hyperref}
\usepackage{url}          
\usepackage{booktabs}  
\usepackage{multirow}
\usepackage{amsfonts} 
\usepackage{nicefrac}       
\usepackage{microtype}
\usepackage{graphicx}
\usepackage[noibid,backend=biber, url=false, isbn=false,urlstamp=false, authordate,giveninits=true,uniquename=mininit,natbib]{biblatex-chicago}

\usepackage{doi}

\title{Measuring the influence of beliefs in belief networks}
\author{ Aleksandar Toma\v{s}evi\'c \\
	Department of Sociology\\
	University of Novi Sad\\
	Novi Sad, Republic of Serbia\\
	\texttt{atomashevic@ff.uns.ac.rs}
}

\renewcommand{\shorttitle}{Measuring the influence of beliefs}

\hypersetup{colorlinks,citecolor = blue,urlcolor=blue,
pdftitle=Measuring the influence of beliefs in belief networks,
pdfauthor={Aleksandar Toma\v{s}evi\'c},
pdfkeywords={european social survey, belief system, network analysis, political networks},
}

\begin{document}

\maketitle

\begin{abstract}

This research proposes a new method for measuring the influence of political beliefs within larger context of belief system networks, based on the advances in psychometric network methods. Using European Social Survey data, we demonstrate this approach on a belief network expressing support for the regime in 29 European countries. Our results show that the influence of beliefs can be related to the consistency and connectivity of the network and that the influence of specific beliefs on a country level has a negative correlation with external indicators from the same domain, suggesting that influential beliefs are related to pressing political issues. These findings suggest that network-based belief influence metrics can be used a new indicator in comparative political research.
\end{abstract}

\section{Introduction}

\section{Introduction}

Belief systems are one of the central topics in scientific study of politics. In the most broader sense, belief systems describe interrelationships of various beliefs related to politics and classical definitions emphasize that those beliefs are bound together by some form of functional interdependence \citep{converse1964nature}. Describing the structure of belief systems, their formal properties and the influential elements of that structure is important for understanding how people draw on different elements of a belief system when evaluating politicians, policies and governments \citep{brandtEvaluatingBeliefSystem2021}. As belief systems may encompass a large number of connected elements, studies seek to identify the most influential or central elements of belief systems \citep{brandtWhatCentralPolitical2019} as those beliefs may be crucial for our understanding of how people reason about political issues, make political decisions and form their knowledge about "what goes with that" in world of politics \citep{hatemiGiveMeAttitudes2016,freederImportanceKnowingWhat2019}.

Recently, a new approach to the study of belief systems emerged, inspired by various methods and tools from network science \citep{dalegeFormalizedAccountAttitudes2016a,boutylineBeliefNetworkAnalysis2017a,brandtWhatCentralPolitical2019,brandtWhatCentralPolitical2019}. From network perspective, measuring influential beliefs is identical to measuring node influence in a network. Influential nodes have a critical role in the spread and propagation of influence in many types of social and physical networks and identification of such nodes has been widely studied by scientists from different disciplines \citep{salavatyIntegratedValueInfluence2020,majiInfluentialSpreadersIdentification2020a,qiuIdentifyingInfluentialNodes2021,zhaoIdentificationInfluentialNodes2021}.

The aim of the current study is to develop a rigorous approach for measuring the influence of beliefs in belief networks based on extracted structural information. Previous studies relied on several classic centrality measures which have been proven either unstable, unreplicable, or not suitable for cross-country comparisons in different political contexts \citep{bringmannWhatCentralityMeasures2019,brandtWhatCentralPolitical2019,brandtEstimatingExaminingReplicability2020a}. Since our aim is not just the identification of the most influential beliefs, the present study is oriented towards more recent and advanced centrality measures.

In order to apply these measures, we will estimate, analyze, and discuss belief networks using the attitudinal entropy (AE) framework \citep{dalegeAttitudinalEntropyAE2018a,dalegeAccurateBeingNoisy2020a,vandermaasPolarizationIndividualsHierarchical2020}, also known as network theory of attitudes \citep{borsboomNetworkAnalysisMultivariate2021}. This approach is based on models analogous to Ising models for ferromagnetism in statistical physics and represents a rare formal model of attitudes. The central idea is that attitudes are high-level properties emerging from configuration lower-level elements such as beliefs, feelings and behaviours. In order to simplify the terminology, we will follow recent work \citep{galesicIntegratingSocialCognitive} and define belief networks as configurations of attitudinal elements which underpin a general attitude.

More formally, belief networks are networks in which nodes represent belief variables and edges represent couplings or interactions between two beliefs at a functional level. Strength of the couplings determines the probability that beliefs are in consistent states \citep{dalegeChangingBeliefsScientific2021}. A positive coupling between two beliefs indicates that on a population level, if we find a person holding one belief, the likelihood that the same person will hold a second belief is increased. Therefore, we can say that two beliefs have influence over one another, but also remain agnostic about the mechanism of causal effects driving this influence \citep{brandtEvaluatingBeliefSystem2021}.

Using AE framework to model belief networks gives us two important formal properties: energy and temperature. Energy of the network can be understood as formalization of the concept of potential dissonance. Potential dissonance refers to the actual inconsistency between beliefs which is translated into felt dissonance if enough attention is given \citep{dalegeChangingBeliefsScientific2021}. Stable, consistent belief networks have lower energy. Probability that the state of any given belief will be in accordance with other beliefs in the network depends on how much attention is focused on the attitude object. Attention therefore represents the strength of the interdependence between beliefs and it has an analogous effect on the network as inverse temperature has on thermodynamic systems. Lower temperature corresponds to a state in which higher attention is given to the attitude object \citep{dalegeAttitudinalEntropyAE2018a,vandermaasPolarizationIndividualsHierarchical2020,dalegeChangingBeliefsScientific2021}.

These formal properties of belief networks are important as we do not expect to find the same distribution of node influence in belief networks with different levels of consistency and different temperatures. To test this assumption, we will estimate multiple belief networks for different social groups for which we expect different levels of energy and temperature and examine if the average belief influence is associated with these formal properties of belief networks between groups. This association is crucial for the validity of influence measurement, because once it is confirmed, further attempts to explain why certain beliefs are influential (or not) in a specific belief system may be based on the notions of dissonance, consistency, and attention of that system. This is especially relevant to the issue of belief or attitude  change, where influential beliefs may be targets of interventions, and recent work has established that change is more likely in high-energy networks \citep{dalegeChangingBeliefsScientific2021}.

Similarly, we aim to prove the external validity of influence measurement by comparing the influence of a specific belief in a belief network with an external measurement of a social issue related to that belief. For example, we will estimate belief networks on subsamples of respondents from different countries, measure the influence of belief related to the satisfaction with democracy in each country and compare these measures with external measures of quality of democracy. We expect a significant correlation between these measures, which would show that belief influence is not a result of some arbitrary network centrality measurement process.

To summarize, our main aim is to develop an approach of measuring the influence of beliefs in belief networks which enables us to associate the influence of a belief with both formal attributes of belief networks and external measurements related to the social issues belonging to the same domain as beliefs whose influence was measured. We will present this approach using an example based on recent European Social Survey data concerning the support of European citizens for regimes in their countries.

\section{Motivating example}

Our motivating example refers to the citizens` attitude towards the regime in their countries. This type of complex attitude has been regarded as crucial for the political legitimacy of liberal democracy through public evaluations of regime performance \citep{norrisDoesDemocraticSatisfaction2012a} and it can be shaped by various dispositions and beliefs \citep{vanryzinPiecesPuzzleLinking2007}. We rely on a multidimensional concept of political support \citep{norrisCriticalCitizensGlobal1999,lindeSatisfactionDemocracyNote2003} which encompasses five  levels or objects of support, four of which can be related to the support for a regime or government and one (political community) which is more general than any specific regime and will not be included in our example. Four regime-related levels are: regime performance, regime institutions, regime principles, and political actors. The macrostate of the belief network we are investigating is the global evaluation of support for the regime which reflects satisfaction with governmental political institutions, principles and specific actors.

Support for regime performance is reflected in satisfaction with various areas of governmental activity (economy, healthcare, democracy) and can be viewed as the belief of the citizens that the regime has fulfilled their expectations and wishes \citep{vanryzinPiecesPuzzleLinking2007,weberTrustPoliticiansSatisfaction2017}. These beliefs emerge as summative satisfaction judgments based on different individual performance perceptions (of various institutions or actors) \citep{vanryzinPiecesPuzzleLinking2007}.

Support for regime institutions is very closely related to political trust in those institutions. This relationship between trust and support is an important issue in political science as both constructs have implications for the political health of a country \citep{warrenTrustDemocracy2018,newtonSocialPoliticalTrust2018}.  Political trust is a type of future-based belief that politicians will serve the citizen's interests and implement their campaign promises \citep{weberTrustPoliticiansSatisfaction2017,taberyNetworkStructureTrust2020}. Recent approaches to political trust from a belief network perspective \citep{taberyNetworkStructureTrust2020,zhangNetworkAnalysisGlobal2021} have unveiled a complex structure of different trust beliefs and different studies had confirmed the association between political trust and beliefs related to satisfaction with democracy and other political institutions \citep{hoboltCitizenSatisfactionDemocracy2012,desimoneExpectationsFutureEconomic2021}. Based on these findings, we include measures of political trust in our belief network as beliefs expressing support for regime institutions.

Support for regime principles in democratic societies is primarily reflected in satisfaction with democracy, which is often viewed as one of the most important elements for the success of a democratic system to implement effective public policies \citep{changMediaUseSatisfaction2018,desimoneExpectationsFutureEconomic2021} and important element of political health of democratic society \citep{claassenDoesPublicSupport2020,valgardssonWhatSatisfactionDemocracy2021}. Satisfaction with democracy has a close relationship with various political performance evaluations which rely on regime's record of policy outputs or outcomes \citep{norrisDoesDemocraticSatisfaction2012a,kollnWhatExplainsDynamics2021}. In order to differentiate between different principles of democratic regimes, we also include several measures related to the intrinsic quality of democratic governance \citep{andrainPoliticalDemocracyTrust2006} such as beliefs related to issue of giving voice to citizens, which is a measure of government responsiveness \citep{sarisAlternativeMeasurementProcedures2009}, and beliefs related to transparency, fairness and impartiality of political procedures.

Finally, support for political actors will be operationalised as trust in politicians and political parties. Taken together, mentioned beliefs will be the constitutive elements of belief networks, which will be used as an example for measuring the influence of specific beliefs.

To produce a more fine-grained analysis, we will consider belief networks of different groups of European citizens. Since we seek to establish the link between the influence of beliefs in belief networks and the formal properties of belief networks (temperature and energy), we will consider belief networks of groups for which we expect different levels of these properties. Therefore, we will divide our sample into subsamples based on: ideological position of the respondent, interest in politics, and the average time spent following news about politics. 

Given that ideology organizes attitudes into thematically consistent networks, has a function of moderating different beliefs and attitudes and given the fact that political issues are considered in light of their ideological implications \citep{jostPoliticalIdeologyIts2009,kayIdeologicalToolboxIdeologies2012}, we can expect different levels of both temperature and energy in groups which are leaning further from the political center, as well as differences in belief influence. In a similar fashion, we can expect that for different levels of political interest and media use, we can expect belief networks with different properties as previous studies have demonstrated strong links between both media use and interest in politics and various political attitudes \citep{strombackMediaMalaiseVirtuous2010,strombackDynamicsPoliticalInterest2013,bimberDigitalMediaPolitical2015}. 

Studies have also shown that support towards the regime in comparative perspective is highly sensitive to different institutional contexts \citep{lindeSatisfactionDemocracyNote2003,desimoneExpectationsFutureEconomic2021}. Therefore, by estimating belief networks on subsamples from different countries, we can expect different structure, energy and temperature levels as well as distribution of important beliefs. As stated in the introduction, important beliefs in each country should be associated with external indicator from the same domain (economy, democracy, healthcare, etc.). When inspecting the attitude towards the regime, we expect that the influential beliefs will be those related to pressing social issues or the domains of weak regime performance.

\section{Measuring node influence}

In terms of centrality metrics on social networks, the notion of influence is generally used to represent the node's ability to affect other nodes and is closely connected to important concepts in social sciences such as power, status or prestige. More generally, we can say that node centrality determines its influence in terms of connectivity, communicability, and controllability \citep{wanSurveyCentralityMetrics2021}. However, before applying the specific centrality metric it is necessary to clarify what node centrality or influence means and represents in a particular context of application \citep{bringmannWhatCentralityMeasures2019}. In classical belief systems literature, the influence or centrality of belief  was referred to as the role beliefs play in the belief system as a whole \citep{converse1964nature}.  In network terms, this means that influential beliefs are driving force of constraint and consistency of the network as they are strongly coupled with their neighboring nodes and they have the ability to directly or indirectly constrain the values of different beliefs throughout the network. However, having many influential beliefs in a network makes consistent states of the network less likely as mutual constraints impose more narrow choice of consistent belief configurations.

Although many new centrality metrics have developed in the last decade, not many of them have been used in the existing applications in network science in general \citep{wanSurveyCentralityMetrics2021}, and particularly in belief networks literature, where strength, betweenness and closeness centrality are almost exclusively used, including some of the specific variations of those measures such as network loadings \citep{christensenEquivalencyFactorNetwork2021}. In biological measurements of node influence, it was shown that single simple centrality measure cannot identify influential nodes in a statistically significant way \citep{delrioHowIdentifyEssential2009}. Therefore, if we want to measure influence with a signle metrics, it needs to capture both local and global information.

K-shell is one of the most important node ranking methods used in social network analysis which is based on both types of information \citep{majiSystematicSurveyInfluential2020}. It describes node's influence starting from node's local information (degree) and iteratively incorporates the global network information \citep{kitsakIdentificationInfluentialSpreaders2010}. The basic idea of this approach is that nodes which are positioned in the core of the network have higher spreading capabilities than those with high node centrality as measured with the classic network centrality measures. The key idea of this approach is to assign an integer value, called k-shell value, which would be representative of node's position within the network in relation to its connectivity patterns. 

The initial step of the k-shell algorithm is to remove all nodes in the network whose degree $k = 1$ and keep removing nodes whose degree $k \leq 1$ after accounting for the updated degrees of nodes given the previous removals. This process continues until there are no nodes in the network with degree $k \leq 1$. All the removed nodes at this point are being assigned with the k-shell value 1. Then the procedure is repeated with incremented k-shell values of 2, 3, and so on. A node with a higher k-shell value is located in a more central position in the network.

One of the biggest strengths of k-shell approach is its robustness in the face of incomplete network information. The k-shell ranking remains robust even when 50 \% of edges are removed from the network \citep{kitsakIdentificationInfluentialSpreaders2010}. However, major drawback of the classic k-shell decomposition approach is that there are many nodes sharing the same k-shell values, so different approaches are proposed in order to increase the monotonicity of k-shell value rankings \citep[see][for a systematic review of k-shell methods]{majiSystematicSurveyInfluential2020}. Central question in attempts to improve k-shell method is how to weight the contribution of the neighborhood k-shell values to node's influence. 

Gravity index centrality (GIC) \citep{maIdentifyingInfluentialSpreaders2016} was proposed as an answer to this problem. This measure takes inspiration from classical gravity formula proposed by Newton and estimates the strength of influence nodes have on each other by simultaneously considering the intrinsic influences of the nodes themselves and the distance between them \citep{liIdentifyingInfluentialSpreaders2019}. K-shell values of two nodes are taken as their mass and the shortest path distance between them as the distance for gravity formula. The influence of node $i$ becomes the sum of the force it exerts to nodes in its neighborhood. This way, nodes with the same k-shell value may have different influence because they are placed in different neighborhoods and have different distances from their influential neighbors.  This approach can be time consuming and computationally heavy for larger networks, but it performs better than degree centrality or classic k-shell method \citep{majiSystematicSurveyInfluential2020}.

\section{Data}

The main analysis of our work is based on European Social Survey  round 9 data \citep{ericEuropeanSocialSurvey2019}. Round 9 is the latest round of the ESS and introduces new rotating module "Justice and fairness in Europe". This module contains new items related to regime principles which are not present in the previous rounds of the survey and therefore we limit our analysis only to this round. We used unweighted data which encompasses 29 countries.
ESS items included in our study are presented in table \ref{tab:ess-data}.

\begin{table}[h!]
\resizebox{\textwidth}{!}{%
\begin{tabular}{@{}clc@{}}
\toprule
\textit{Name} & \multicolumn{1}{c}{\textit{Survey item}}                                                    & \textit{Group}      \\ \midrule
Economy       & On the whole how satisfied are you with the present state of the economy in [country]?      & Regime performance  \\
Education     & Please say what you think overall about the state of education in [country] nowadays?       & Regime performance  \\
Health        & Please say what you think overall about the state of health services in [country] nowadays? & Regime performance  \\
Democracy     & And on the whole, how satisfied are you with the way democracy works in [country]?          & Regime principles   \\
PeopleAllow & How much would you say the political system in [country] allows people like you to have a say in what the government does?        & Regime principles \\
PeopleInf   & And how much would you say that the political system in [country] allows people like you to have an influence on politics?        & Regime principles \\
Transparent & How much would you say that decisions in [country] politics are transparent, meaning that everyone can see how they were made?    & Regime principles \\
CitInterest & How much would you say that the government in [country] takes into account the interests of all citizens?                         & Regime principles \\
FairChance  & How much would you say that the political system in [country] ensures that everyone has a fair chance to participate in politics? & Regime principles \\
Parliament    & How much you personally trust [country]'s parliament?                                       & Regime institutions \\
Police        & How much you personally trust the police                                                    & Regime institutions \\
Legal         & How much you personally trust the legal system?                                             & Regime institutions \\
Politicians   & How much you personally trust politicians?                                                  & Political figures   \\
Parties       & How much you personally trust political parties?                                            & Political figures   \\
News        & On a typical day, about how much time do you spend watching, reading or listening to news about politics and current affairs?     & Grouping variable \\
PolInt        & How interested would you say you are in politics?                                           & Grouping variable   \\
LRScale     & In politics people sometimes talk of 'left' and 'right'. Where would you place yourself on this scale?                            & Grouping variable \\ \bottomrule
\end{tabular}%
}
\caption{European Social Survey items, variable names and groups}
\label{tab:ess-data}
\end{table}

Following previous work \citep{dalegeChangingBeliefsScientific2021}, we transformed all belief variables to 0-1 scale. For variable related to following news about politics, we transformed the original variable and created four groups based on quartile values in minutes (0-30, 30-60, 60-90, 90+). Finally, for the ideological left-right scale we transformed the original variable and created five groups (Far Left, Left,Center, Right and Far Right).

External country-level indicators are gather from public data sets from different sources (see Online Supplement for more information). Since ESS data contain survey items recorded in late 2018 and early 2019, country-level indicators should describe the quality of government and regime performance at the time or shortly preceding the time at which the survey responses were recorded.

\section{Methods}

In the previous sections, we have described the motivating example for our study and shown the ESS variables we will use for the analysis. Before, we proceed with the estimation of networks, we will employ Unique variable analysis (UVA) \citep{christensenUniqueVariableAnalysis2020} in order to check if there are redundancies between the survey items we have chosen. If such redundancies are detected we will remove problematic items to achieve a more parsimonious representation of belief networks. 

The next step is to estimate belief networks in different subsamples. This will be achieved using psychometric network analysis methods \citep{epskampGeneralizedNetworkPsychometrics2017,epskampEstimatingPsychologicalNetworks2018,borsboomNetworkAnalysisMultivariate2021} with partial correlations representing couplings between beliefs. Finally, for each estimated belief network we will measure the influence of beliefs using GIC \citep{maIdentifyingInfluentialSpreaders2016}.

The entire analysis was performed in R \citep{baser}. European Social Survey data was imported using essurvey package \citep{essurvey}. Main analysis was based on EGAnet \citep{golino2021eganet}, psychonetrics \citep{psychonetrics}, qgraph \citep{qgraph}. We provide new R implementation of weighted k-shell method and GIC based on igraph \citep{igraph} package. The entirety of data, code, notebooks and results can be found in an online repository: \href{https://github.com/atomashevic/belief-influence}{https://github.com/atomashevic/belief-influence}.

\subsection{Unique Variable Analysis}

Unique variable analysis (UVA) is a novel approach for detecting and removing redundant survey items from the analysis \citep{christensenUniqueVariableAnalysis2020}. Because elements of belief networks are assumed to be casually autonomous, redundancy of a pair of survey items due to their shared causes problems for the interpretation of the entire network. Nodes who are mutually redundant are strongly connected and have an identical or quite similar global network position, thus negatively affecting almost every approach to measure the influence of the nodes as it would lead to inflated measurements for those nodes having one or more redundant pairs.

Our application of UVA is based on weighted topological overlap (wTO) \citep{zhangGeneralFrameworkWeighted2005}. Weighted topological overlap is a network measure of redundancy which quantifies the similarity of two nodes' pattern of connections with all other nodes in the network, including the similarities in the weights of edges they have with common nearest neighbors. The basic idea of UVA is to estimate wTO for each pair of variables and values above a a chosen threshold used to evaluate which groups of nodes are considered highly redundant and will be removed from the network.

This is a simple approach which does not require estimation of a latent factor model to detect redundant variables, and no references are made to the factor or latent structure of the variable set being investigated. In order to estimate wTO, UVA method relies on estimating the network model using the standard approach in the psychological network literature, partial correlation network estimation with graphical lasso regularisation with EBIC model selection and tuning parameter $\gamma=0.5$ \citep{epskampEstimatingPsychologicalNetworks2018,golinoExploratoryGraphAnalysis2017}. For wTO, we applied the threshold of 0.25 which is the mean value of previously reported evidence for wTO cutoff values \citep{nowickDifferencesHumanChimpanzee2009,christensenUniqueVariableAnalysis2020}. Once groups of redundant variables (pairs or triads) are identified, items which relate to a broader political concept and have higher standard deviation remain in network while others are removed.

\subsection{Network Structure Estimation}

Estimation of the belief network relies on the methods belonging to the psychometric network analysis \citep{,epskampGeneralizedNetworkPsychometrics2017,marsmanIntroductionNetworkPsychometrics2018a,borsboomNetworkAnalysisMultivariate2021} or the analysis of multivariate psychometric data using network structure estimation and network description in order to capture complex interactions between variables. Using this approach, the joint probability distribution of a set of variables is described in terms of pairwise statistical interactions. This results in a graphical model knows as pairwise Markov random field (PMRF) \citep{lauritzen1996graphical,epskampPsychometricNetworkModels,borsboomNetworkAnalysisMultivariate2021}. If we view this type of a graphical model as a network than each variable is represented as a network node and nodes are connected by edges. Edge between two nodes represents a conditional association between two variables, conditional on all other variables in the data. For multivariate normal data, these associations can are represented by partial correlation coefficients who are then directly used as edge weights in the resulting network. If data are continuous, as in our case, resulting is PMRF is the Gaussian Graphical model (GGM). Formally, GGM is a model for a variance covariance matrix $\mathbf{\Sigma}$:

\begin{equation}
    \label{eq:ggm}
    \Sigma=\Delta(I-\Omega)^{-1} \Delta,
\end{equation}

where $\Delta$ is a diagonal scaling matrix, $I$ is the identity matrix, and $\Omega$ is a square symmetrical model matrix with zeroes on the diagonal and partial correlation coefficients on the off diagonal (the edge weights of the network).

In practice, we will use an automated model selection using step-up search procedure. This procedure starts from an empty network and iteratively add edges between nodes in a manner which increases the network model's fit to the data as estimated by Bayesian Information Criterion (BIC). New edges are partial correlations whose addition to the model has the largest value of statistically significant modification index. This approach is preferred for smaller networks where we do not expect sparse structure \citep{borsboomNetworkAnalysisMultivariate2021} rather than regularized approaches which are more common and come with the assumption of network sparsity. Regularized approach was used only in the previous step (UVA), where the goal was not to approximate the network structure but rather to explore the dyadic redundancies between beliefs.

For every analysis that includes multiple groups, we perform model selection for different classes of models. Given the GGM model shown in equation \ref{eq:ggm}, we can constrain any model parameters to be strictly equal or free between groups. Of special importance are the models where we constrain the network structure itself $\Omega$ and/or and the scaling matrix $\Delta$ to be equal between groups. The relationship of these constrained models to models without constraints will provide information about whether networks and their formal properties differ between groups or not.

\subsection{Formal properties of belief networks}

As we stated in the introduction, the two formal properties of belief networks we are interested in are energy and temperature. We modify the previous approach \citep{dalegeAttitudinalEntropyAE2018a,dalegeChangingBeliefsScientific2021} and first estimate the energy for the individual survey respondent $k$:

\begin{equation}
    \label{eq:energy}
    H_k = - \sum_{i,j} w_{ij} b_{i_k}b_{j_k},
\end{equation}

where $\omega_{i,j}$ is the element of $\Omega$ partial correlation coefficient between pair of beliefs $b_i$ and $b_j$, and $b_{i_k}$ and $b_{j_k}$ are the values of two belief variables for respondent $k$. The average energy $\hat{H}$ of the belief network is the mean value of all individual energies.

For temperature, we use the approach of the previous studies \citep{dalegeAttitudinalEntropyAE2018a,dalegeChangingBeliefsScientific2021} and calculate it as the average value of the elements of scaling matrix $\Delta$. Lower average scaling value indicates higher overall partial correlations in the network.  Correlations implied by the chosen model result from dividing the measured partial correlations between two beliefs by the product of their scaling values. Therefore, the average scaling value approximates the temperature of the belief network.

\subsection{Node influence}

We applied the k-shell method for weighted networks \citep{garasShellDecompositionMethod2012} in order to retrieve k-shell values for nodes in our networks. The core of this method is the alternative concept of node degree, which is calculated using its unweighted degree (number of edges) and the sum of edge weights (also known as node strength). The weighted degree of node $i$ is:

\begin{equation}
\label{eq:wdegree}
k_{i}' = [ k_i^{\alpha}(\sum_{j}^{k_j} w_{ij})^{\beta} ]^{\frac{1}{\alpha + \beta}},
\end{equation}

where $k_i$ is unweighted degree and $\sum_{j}^{k_j} w_{ij}$ is the sum of edge weights (strength centrality). In case of $\alpha = \beta = 1$ the weights and the degree are treated equally and weighted degree is a geometric mean of degree and strength centrality. In our case, we expect dense networks with few nodes and therefore to increase to monotonicity of this measure we set $\alpha = 1$ and $\beta =2$ to put more emphasis on edge weights.

Once we have the weighted degrees, we apply the original k-shell decomposition algorithm \citep{kitsakIdentificationInfluentialSpreaders2010} to obtain k-shell values $k_s$. Then we can proceed to calculate GIC centrality \citep{maIdentifyingInfluentialSpreaders2016} for any node $i$:

\begin{equation}
    \label{eq:gic}
    GIC(i) = \sum_{j \in \theta_i} \frac{ks_i \times ks_j}{dist_{ij}^2},
\end{equation}

where $\theta_i$ is the nearest neighborhood of node $i$, $ks_i$ and $ks_j$ are k-shell values and $dist_{ij}^2$ is the squared shortest path distance obtained using Dijkstra algorithm. Shortest path distances were calculated using the original degrees and not weighted ones. Weighted degrees are used only for k-shell decomposition and discarded afterwards.

\section{Results}

The first step of the analysis was the application of Unique Variable analysis to remove the redundant variables. Before the first round of UVA, the highest value of wTO and the highest redundancy was between variables  capturing trust in the legal system and trust in police (0.943). After first round of UVA, the following variables were removed: \textit{Transparent, FairChance, PeopleInf, Police, Parties}. This removal resulted with significantly smaller wTO and highest redundancy after 1st round of UVA was between trust in Parliament and trust in legal system (0.352). We performed a second round UVA and removed following variables from redundant pairs with wTO higher than the threshold (0.25): \textit{Legal, Politicians}.

After removing the redundant variables, a set of 7 variables remained: \textit{Economy, Education, Health, Democracy, PeopleAllow, CitInterest and Parliament}. For this set of variables we estimated a set of belief network models for several groups of respondents. These groups are based on variables \textit{News, PolInt and LRScale} and in total we estimated belief networks for 13 groups. 
For each grouping variable, we constructed 8 statistical network models. For benchmark purposes, we created 4 empty network models and 4 nonempty models using step up procedure with different constraint conditions between different groups: all parameters free, equal temperature, equal edges and weights, equal temperature, weights and temperature.

For all cases, model with no constraints between the groups had the lowest values of BIC and therefore was selected as best fitting model. Second best model is the model with equal edges and weights. This means that in all cases the estimated network performed better than the null model and that there are differences in network structure and temperature between various groups. 

\begin{figure}[h!]
    \centering
    \includegraphics[width=0.9\textwidth]{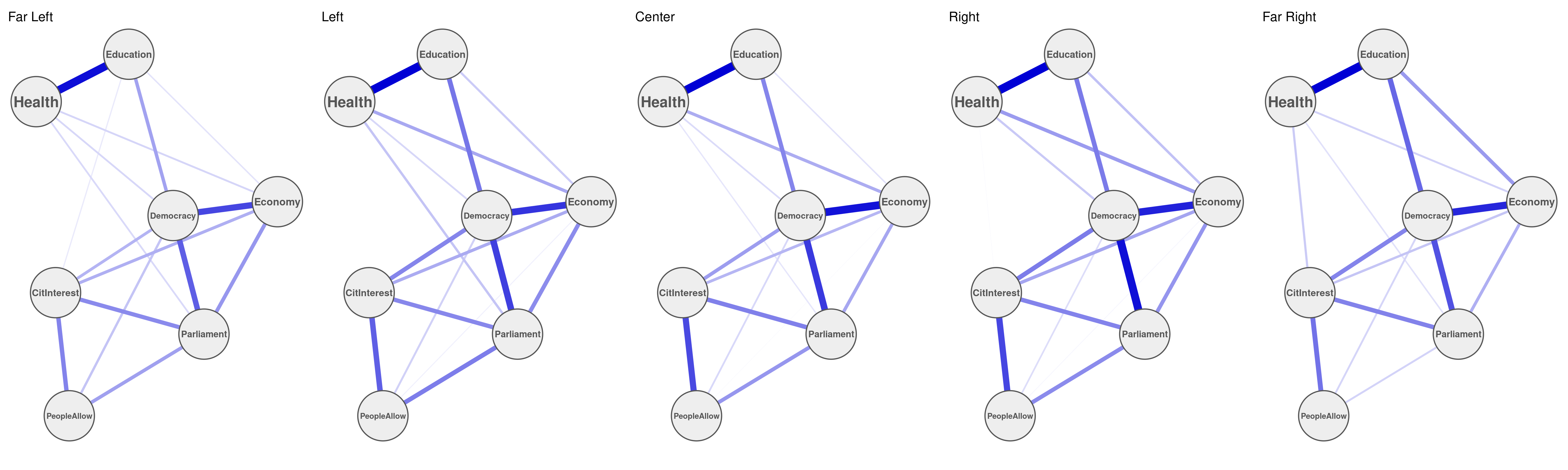}
    \caption{Belief networks of different ideological groups}
    \label{fig:bn-lr}
\end{figure}

For groups estimated on the basis of the ideological self-identification of the respondent, differences between belief networks can be seen in figure \ref{fig:bn-lr}. All 5 belief networks exhibit low dimensionality and a relatively simple psychological structure which is expected in belief systems \citep{endersConspiracyBeliefsForm2021}.

\begin{figure}[h!]
    \centering
    \includegraphics[width=0.9\textwidth]{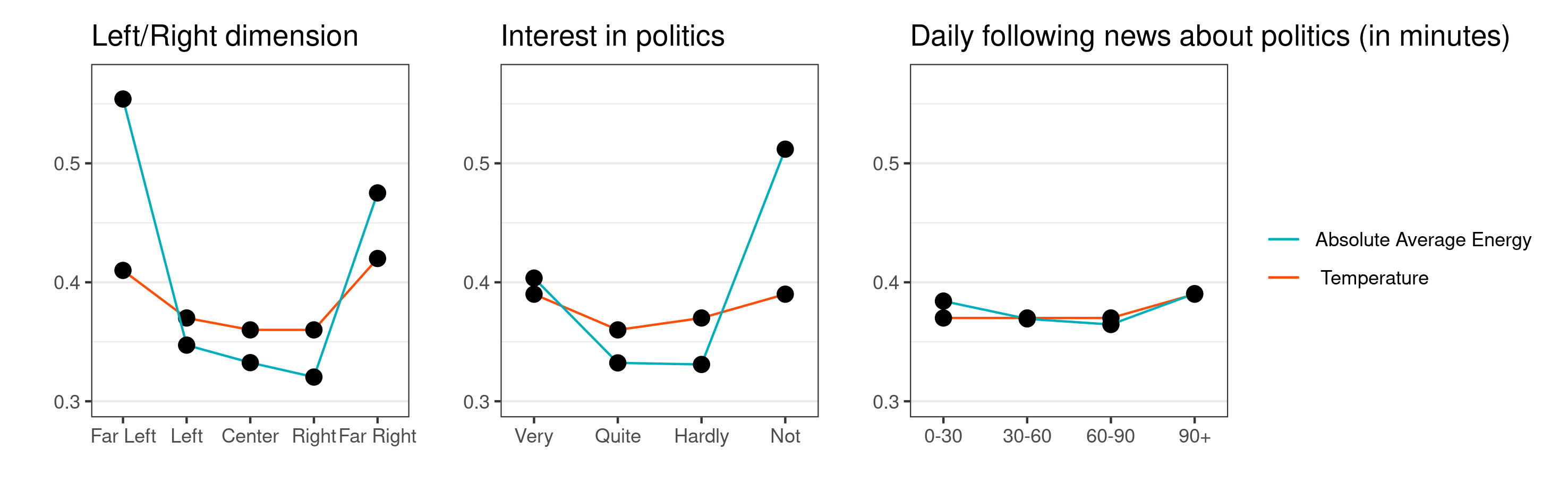}
    \caption{Differences in absolute average energy and temperature for different groups of respondents. \textit{Note:} Higher values of absolute energy correspond to lower energy (consistent belief network)}
    \label{fig:en-temp}
\end{figure}

In these belief networks, some edges (e.g. Health-Education) are constant between groups, while others differ (e.g. Democracy-Parliament). Given the simple network structure, these slight differences in edges and their weights produce differences in belief network energy and temperature, which are shown on figure \ref{fig:en-temp}.

We see that groups on the extremes of ideology, interest in politics, and news watching have consistently higher absolute energy and more consistent belief networks. In the case of ideology, we see that respondents belonging to Far Left have the most consistent attitude towards the regime, followed by Far Right. Similar trend was observed for temperature, with groups on the scale extremes having higher temperature meaning belief system has weaker 
connectivity and that less attention is payed to the attitude object. This result is interesting in the case of ideology groups as it seems that respondents on ideological extremes pay less attention to the activities and performance of the regime. This finding supports the idea that ideology operates as a cognitive heuristic and reduces the attention needed for an individual to form a consistent attitude towards the attitude object \citep{rudolphPoliticalTrustIdeology2005}.

\begin{figure}[h!]
    \centering
    \includegraphics[width=0.55\textwidth]{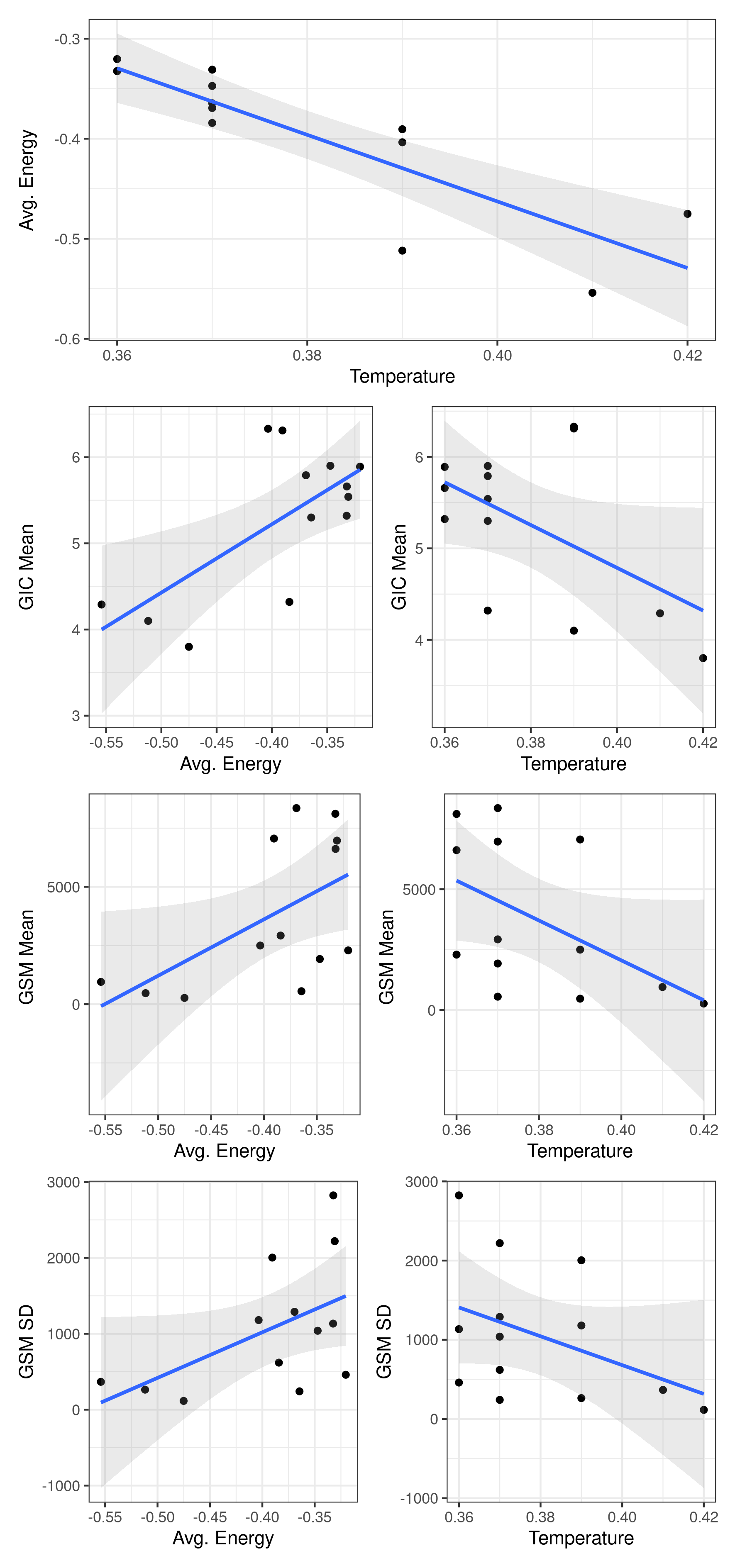}
    \caption{Scatterplots showing relationships of formal network properties and GIC. \textit{First row:} Belief network temperature and average energy. \textit{Second row, left:} Average energy and average GIC of belief network. \textit{Second row, right:} Temperature and average GIC of belief network.}
    \label{fig:scatter-01}
\end{figure}

 In most groups, satisfaction with democracy (SWD) is the most influential node (results are provided in online supplementary material). As stated before, we are more interested in the relationships between influence measure and formal properties of belief networks. They are shown on figure \ref{fig:scatter-01}. There exists a clear inverse linear relationship between energy and temperature. Consistent low-energy belief networks have higher temperature, which means lower connectivity and attention. These results point towards a trade-off in this type of belief networks, where either consistency in attitude towards the regime is obtained (on average in a group) or higher state of belief connectivity and higher attention to the attitude object is obtained.

The second row of panels on figure \ref{fig:scatter-01} shows the relationship of the mean value of GIC with the average energy and temperature of the belief network. In low-energy highly consistent networks, we found a low average influence of beliefs and it seems that higher consistency of beliefs is possible in the absence of highly influential beliefs, while the reverse is found in case of temperature.

\begin{figure}[hbt!]
	\centering
	\includegraphics[width=0.55\textwidth]{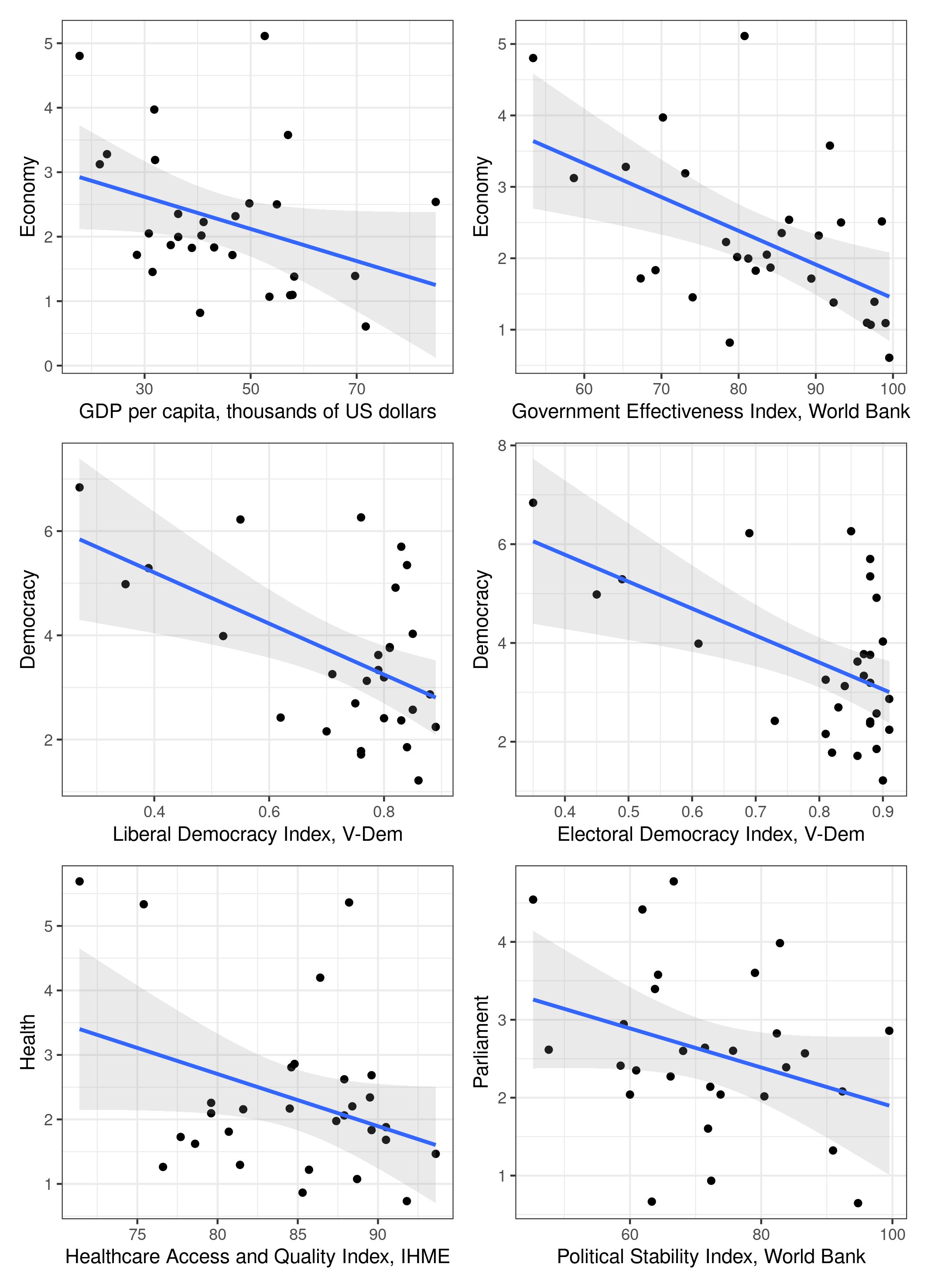}
	\caption{Scatterplots showing the relationship between belief influence (measured by GIC) and country-level external indicators}
	\label{fig:scatter-02}
\end{figure}

Finally, the relationship between the influence of beliefs as measured by GIC and external country-level measures associated with the corresponding domains is shown on figure \ref{fig:scatter-02}. The influence of belief expressing satisfaction with the national economy is negatively correlated with GDP per capita ($r=-0.352$, $p=0.06$) and with Government Effectiveness index ($r=-0.530$, $p=0.03$) which is related to the quality of public services and national policy formulation and implementation, including policies related to the national economy \citep{kaufmannWorldwideGovernanceIndicators2010}.

Similar results are obtained for influence of satisfaction with democracy and V-Dem indices of Liberal ($r=-0.52$, $p=0.004$) and Electoral democracy ($r=-0.52$, $p=0.003$). As the majority of European countries in the sample score high on these indices, it is important that we observe in those countries with lower score a higher influence of satisfaction with democracy. The influence of satisfaction with healthcare system is also found to be greater in countries with lower values of Healthcare Access and Quality Index ($r=-0.344$, $p=0.07$). Similar correlation is found between the influence of trust in national parliament and Political stability index, which indicates the likelihood that the government will be destabilized or overthrown by unconstitutional or violent means ($r=-0.31$, $p=0.09$).

\section{Limitations}

Study has several notable limitations. First, the analysis is based on ESS data, which means a fixed, closed set of survey items are available and form the basis of the belief system analysis. If different survey items were available, after redundancy analysis, the resulting belief network may have different number of different items, although UVA minimizes this risk as we reduced the number of items from a larger set of items present in the survey to core questions with minimal redundancy.

Second, we use the unweighted, raw version of the survey data. Although we are not interested in country-level means of survey items, structural post-stratification inequalities in national subsamples may influence the results concerning comparisons of networks estimated from those subsamples.

Third, influence measures are selected based on literature review and the analysis of previous applications on network different than belief networks. It is possible that there is a another subset of all proposed centrality measures capable of measuring the influence of beliefs in this context and not taken into consideration in our paper.

Fourth, reported correlations are based on 29 country-level subsamples and reported $p$-values have to be treated with caution especially given the observational nature of the study.

\section{Discussion}

Main findings of our study contribute to the hardcore of theory of belief system networks \citep[see][]{brandtEvaluatingBeliefSystem2021}. Our results present evidence of the relationship between previously established properties of belief networks and new measures of belief influence. Also, we have shown that similar relationship exists between belief influence and external indicators related to same belief domain. Taken together, these results support the claim that centrality-based influence metrics in belief networks provide valuable information, in contrast to the studies questioning the utility of node centrality in psychometric network analysis. \citep{bringmannWhatCentralityMeasures2019}.

Previous results in network theory of attitudes have shown that when attitude is important, we have higher consistency and stability of the attitude, because different attitude elements rein each other in systems with low temperature \citep{borsboomNetworkAnalysisMultivariate2021}. Our results show that belief networks with higher average influence have lower temperature. Higher average influence implies higher connectivity and higher average centrality in the network as it is highly probable that on the neighborhood level, nodes are capable of influencing on another.

We find higher consistency of attitude towards the regime in groups positioned on ideological extremes, but the connectivity of these networks is lower. In this case, ideology provides consistency but without strong links between non-ideological beliefs. This finding is in line with recent study \citep{blankenshipSelfValidatingRolePolitical2021} which demonstrates that ideology can serve as a mechanism to validate one's reactions to different political issues and increase certainty in thoughts related to them. If the ideology a the proxy mechanism which provides certainty in our belief networks, then it would be expected that attention towards the attitude object (the regime) becomes lower in groups positioned on ideological extremes.

Further studies are needed to demonstrate that our approach is viable for different belief networks and GIC's performance needs to be evaluated in comparison with both simple and advanced centrality metrics. Psychometric network analysis and the theory of belief system networks have a lot to offer to comparative political science studies and network-based belief influence measurement may be the right first step towards bringing these fields closer through multidisciplinary collaborations.

\printbibliography

\end{document}